# Overcoming Bifurcation Instability in High-Repetition-Rate Ho:YLF Regenerative Amplifiers

PETER KROETZ[1,2,*], AXEL RUEHL[3], GOURAB CHATTERJEE[1,2], ANNE-LAURE CALENDRON[1,4,5], KRISHNA MURARI[1,3,4], HUSEYIN CANKAYA[1,3,5], PENG LI[3], FRANZ X. KÄRTNER[1,3,4,5], INGMAR HARTL[3], AND R. J. DWAYNE MILLER[1,2,5,6]

[1]Center for Free-Electron Laser Science (CFEL), Notkestraße 85, D-22607 Hamburg, Germany
[2]Max-Planck Institute for the Structure and Dynamics of Matter (MPSD), Luruper Chaussee 149, 22761 Hamburg, Germany
[3]Deutsches Elektronen-Synchrotron (DESY), Notkestrasse 85, 22607 Hamburg, Germany
[4]Department of Physics, University of Hamburg, 22761 Hamburg, Germany
[5]Centre for Ultrafast Imaging (CUI), Universität Hamburg, Luruper Chaussee 149, 22761 Hamburg, Germany
[6]Departments of Chemistry and Physics, University of Toronto, Toronto M5S 1A7, Canada
*Corresponding author: peter.kroetz@mpsd.mpg.de

**We demonstrate a Ho:YLF regenerative amplifier (RA) overcoming bifurcation instability and consequently achieving high extraction energies of 6.9 mJ at a repetition rate of 1 kHz with pulse-to-pulse fluctuations of 1.1%. Measurements of the output pulse energy, corroborated by numerical simulations, identify an operation point that allows high-energy pulse extraction at a minimum noise level. Complete suppression of the onset of bifurcation was achieved by gain saturation after each pumping cycle in the Ho:YLF crystal via lowering the repetition rate and cooling the crystal. Even for moderate cooling, a significant temperature dependence of the Ho:YLF RA performance was observed.**

High-energy ultrashort-pulse laser sources in the mid-infrared have recently garnered attention for their pivotal role in high harmonic generation and strong-field physics [1]. Tunable mid-IR sources based on optical parametric amplification (OPA) may be conveniently realized by pump laser systems at an operational wavelength of 2 μm. This offers a manifold of advantages over the more conventional 1-μm pump sources due to the possibility of using highly efficient non-oxide crystals (for example, ZnGeP2) for the OPA. Of late, promising 2-μm-based femtosecond-seeded regenerative amplifiers (RAs) have been demonstrated, based on the gain materials Ho:YAG, Ho:YLF and Tm:YAP [2-8]. For instance, Ho:YAG RAs have yielded more than 2 mJ at 1 kHz [2] and up to 3 mJ at 5 kHz [3]. Besides, for Ho:YLF, Grafenstein et al. [4] and Dergachev et al. [5] reported output pulse energies up to ∼2 mJ at a repetition rate of 1 kHz, whereas Hemmer et al. [6] demonstrated pulse energies as high as 5.5 mJ, albeit at 100 Hz. In addition, Hemmer et al. also developed a highly efficient cryogenically cooled Ho:YLF single-pass amplifier, yielding up to 39 mJ [6], taking advantage of the strong temperature dependence of the absorption and emission cross-sections of Ho-doped gain materials due to their quasi-three-level nature [6, 9, 10]. Tm:YAP RAs, on the other hand, can generate sub-picosecond pulses (for example, < 400 fs pulses at 0.7 mJ [8]), in contrast to the multi-picosecond pulses from Ho-based RAs.

Of paramount importance for the design of a RA is its susceptibility to output pulse energy fluctuations. This may be attributed to various factors such as pump fluctuations and inherent noise sources [11]. In addition, it is well-known that for repetition rates on the order of the inverse lifetime of the gain medium (or higher), the RA may fluctuate periodically between two or more output energies [12]. The RA is then said to be driven into a bifurcation regime (see Fig. 1 (a)). The underlying cause for bifurcation can be traced to a disturbed equilibrium in the gain for consecutive pump and amplification cycles close to gain saturation (as shown in Fig. 1 (b) for three consecutive cycles). This effect is especially stringent at high repetition rates, where this equilibrium is more difficult to maintain [12, 13]. However, in principle, it should be possible to compensate the adverse effect of the higher repetition rate on the gain by a higher pump power so as to maintain a constant gain in the crystal. This is further illustrated in Figure 1(c), which shows the iso-gain contour as a function of the pump power and the repetition rate (for a given cavity design), separating the regimes with and without susceptibility to bifurcation. It is thus possible to drive a RA at a repetition rate higher than what is allowed by the inverse lifetime of the gain medium, by suitably increasing the pump power so that the system cannot be driven into bifurcation. Due to the different time intervals for gain and amplification, the bifurcation dynamics do not depend on the pulse duration [12].

To the best of our knowledge, all 2-μm RA systems that have been reported so far were either operated at low repetition rates (close to the inverse of the lifetime of the gain medium) [6], thereby completely suppressing the onset of bifurcation, or alternatively at higher repetition rates [7], but only until bifurcation sets in. These RA systems therefore constitute a compromise among achievable pulse energy, repetition rate, and low-noise operation.

In this Letter, we demonstrate the first experimental measurement of a complete bifurcation trajectory in a 2-μm RA including the existence of a 2nd operation point (OP) with the potential of high energy extraction with minimum pulse-to-pulse fluctuations. This 2nd OP is accessible by driving the gain medium close to gain saturation and by carefully choosing an appropriate crystal length, such that despite complete gain depletion, the energy content of the output pulse does not exceed any damage threshold. By operating our Ho:YLF RA at the 2nd OP, we were able to achieve pulse energies of 6.9 mJ at 1 kHz repetition-rate with pulse-to-pulse fluctuations of only 1.1%. This is the highest pulse energy achieved to date, with a RA operating at 2 μm at any repetition rate. We also demonstrate effective bifurcation suppression by gain saturation up to repetition rates of 750 Hz, which exceeds the inverse life-time of Ho:YLF by more than an order of magnitude.

A schematic of the experimental setup is shown in Fig. 2. The gain medium is a TEC-cooled 9-mm long Ho:YLF crystal, with an aperture of 5 mm × 5 mm, and a doping concentration of 1%, yielding a relatively modest CW output power of up to 14.5 W for 18% output coupling, despite an incident pump power of 108 W. The unabsorbed pump power of ~75 W could be used for further amplification stages, to increase overall amplifier efficiency. The RA cavity length is 1.7 m and simulations for the cavity mode show negligible astigmatism. The beam diameter in the Ho:YLF crystal is 1 mm. To minimize the effect of the Pockels cell on the B-integral, a rather large and collimated beam diameter of 2.4 mm is maintained in the Pockels cell.

The seed source is a home-built passively mode-locked Ho:fiber soliton oscillator [14]. Its output pulses exhibit a spectral width of 7 nm (FWHM), centered at 2053 nm. After stretching in a chirped volume Bragg grating (CVBG) with a chirp rate of 41 ps/nm, the oscillator pulses exhibit a 3.25-nm (FWHM) spectral bandwidth (Fig. 3 (c)) and a pulse energy of ~ 0.3 nJ. Based on the specified chirp rate and grating efficiency, the pulses were estimated to have a pulse duration of 533 ps with a total efficiency of 66% in the 4-pass geometry through the CVBG. The input and output coupling of the seed and amplified pulses, respectively, were mediated through a 8-mm aperture Faraday rotator (FastPulse Technology, Inc.) and a 6-mm aperture RTP Pockels cell.

The Ho:YLF RA was pumped by a commercial randomly-polarized thulium-doped fiber laser module with a CW output power of 120 W, centered at 1939.5 nm. The pump laser output was focused by a relay imaging system to a spot size of ~1 mm, resulting in pump intensities of ~30 kW/cm$^2$. The transverse profile of the pump beam output deteriorates severely due to atmospheric absorption. Consequently, purging the laser system with dry nitrogen results in a significant improvement in the stability and the spatial profile of the pump beam. Figures 3 (a) and (b) show a remarkable similarity between the simulated and experimentally measured bifurcation trajectories for our Ho:YLF system, operated at 1 kHz repetition rate. To calculate the measured energy density probability, ~15000 consecutive output pulses were recorded for each round trip number using a commercial calibrated energy-meter. The simulations were performed based on adapted monochromatic Frantz-Nodvik equations [15, 16]. We assume a 11% loss per round trip and a 3% fluctuation in the pump fluence, consistent with the performance specifications of the pump laser. The only free parameters in the simulations were the absorption and emission cross-sections at the pump and the lasing wavelengths, which were optimized to closely reproduce the measurements shown in Fig. 3 (b) and were found to be 50% and 70% of the respective literature values [17]. This deviation could be explained by a higher temperature in the bulk of the crystal, compared to the temperature of the crystal holder. Both the simulations and the experimental measurements identify two distinct OPs that offer stable and high pulse-energy extraction. The 1st OP is identified as the round trip before the noise exceeds 10%, whereas the 2nd OP has the minimum noise. These two OPs are separated by a range of round trips that exhibit bi-stable output pulses even in the absence of noise, consistent with theoretical predictions [12]. The energy spread for a given round trip number can be attributed to the noise in the pump and seed sources, particularly outside the bi-stable regime. We use the standard deviation of this energy spread as a measure of the stability of the output pulses, determined as 1.1% for the 1 kHz RA system at the 2nd OP. Compared to the 1st OP, we observed not only an almost 3-fold increase of the pulse energy but also a 5-fold reduction of the pulse-to-pulse fluctuation. The corresponding optical spectrum measured at the maximum pulse energy, shown in Fig. 3 (c), supports a minimum pulse duration of 1.9 ps, as can be seen from the calculated Fourier-limited pulse profile in Fig. 3 (d).

Fig. 4 (a) shows the output pulse energy versus round trip number for repetition rates of 10 Hz, 500 Hz and 1 kHz. By lowering the repetition rate, we were able to achieve pulse energies of up to 13 mJ at 10 Hz and at 2.5°C crystal temperature, resulting in peak fluences of up to 3.7 J/cm2 in the Ho:YLF crystal. Despite the 100 times higher pump fluence per pumping cycle, when comparing the system at 1 kHz and 10 Hz, the maximum output pulse energy increases only by a factor of 1.9. This is a clear indication of gain saturation in the laser crystal, a precondition for bifurcation suppression. As a consequence, the bifurcation instability of the laser could be completely suppressed for repetition rates of ≤ 750 Hz at the lowest temperature. It is to be noted that this suppression leads to a similar noise performance, at the maximum extractable pulse energy, compared to operation at the 2nd OP. The temperature of the Ho:YLF crystal (measured at the crystal holder) had a significant effect on the system performance, as shown in Fig. 4 (b) for a repetition rate of 750 Hz. The maximum output pulse energy increased by 33% while lowering the crystal temperature from 19 to 2.5°C. Furthermore, the onset of bifurcation was completely suppressed. This suppression effect can be explained with higher absorbed pump powers of Ho:YLF at lower temperatures, leading to an increased gain saturation. For instance, a higher absorption at lower temperatures is supported by the observation that in CW operation, for an incident pump power of 13.2 W, the unabsorbed pump power decreases from 8.1 W to 7.7 W on decreasing the temperature of the crystal holder from 18°C to 2.8°C, while the CW output power increases from 0.82 W to 1.33 W for 11% output coupling.

Most notably, for all measurements across all repetition rates and temperatures, it was observed that the RA exhibits the lowest output pulse noise for a specific value of the round trip number, as the system approaches gain depletion, consistent with simulations [18]. This usually corresponds to the highest output pulse energy.

A summary of the obtained pulse energies as a function of the repetition rate is shown in Fig. 5 (a), together with the pulse energies reported in literature for similar laser systems. The B-integral of the system (also shown in Fig. 5 (a)) was estimated between 1 – 1.4, assuming a 377-ps pulse duration and a spectral width of 2.3 nm. Despite the increase in the pulse energy at lower repetition rates, the B-integral stays almost constant, due to the decrease in the round trip number. Since the value for the non-linear refractive index of the RTP Pockels cell is unknown, we used the value of its isomorph KTP (12×10-16 cm2/W) [19] for the calculation of the B-integral, which has similar linear and non-linear optical properties [20]. Under this assumption, the Pockels cell contributes to approximately 75% of the total B-integral. At the most stable operation point, across all repetition rates, the RA exhibits a pulse-to-pulse fluctuation less than 1.6%.

For the measurement of the initial results presented above, the RA was operated for more than 12 hours before the crystal was damaged in the volume. A thermal influence as damaging mechanism seems likely, since the beam shape of the output pulses deteriorates at higher pump powers as well as at higher repetition rates (see, for example, Fig. 5 (b)). Beam distortions were also observed in CW operation for high intra-cavity powers (low output-coupling losses). We are currently reworking the crystal cooling to facilitate reliable long-term operation of the laser. Pulse compression of the amplified pulses in a CVBG identical to the one used for stretching is ongoing. As in the case of the CVBG stretcher, we expect a compressor efficiency of 66%. Assuming the Fourier-limited pulse duration of 1.9 ps, the laser system supports pulses with 2.4 GW peak power at 1 kHz.

In summary, we have demonstrated the optimized performance of a Ho:YLF RA at 1 kHz repetition rate, overcoming bifurcation instability and obtaining output pulse energies as high as 6.9 mJ at 1 kHz and up to 13 mJ at 10 Hz. These are the highest pulse energies reported to date for a 2-µm RA system at any repetition rate. The most important achievement in terms of utility of this system is the dramatic reduction in noise by exploiting the 2nd OP. The lowest noise operation is related to the gain depletion during the seed amplification process and noise values as low as 1.1% could be achieved at 1 kHz. We have numerically and experimentally demonstrated a complete bifurcation trajectory and we identified the key 2nd OP that exhibits an almost 3-fold increase of the pulse energy and a 5-fold reduction of the pulse-to-pulse fluctuation compared to the 1st OP. We furthermore demonstrated that driving the gain further to saturation – either by reducing the repetition rate or by lowering the temperature of the crystal holder – can lead to a complete suppression of the onset of bifurcation in the repetition rate range of interest. A suppression was achieved for the onset of bifurcation for repetition rates ≤ 750 Hz, more than 10 times the inverse life-time of Ho:YLF. The effect of cooling the Ho:YLF crystals on the system performance is significant. Even for a moderate temperature decrease from 19 °C to 2.5 °C, the output pulse energy increased by 33%. We can therefore envisage a scenario where a more efficient cooling will enable us to extend our current system performance to yet higher extraction energies and even higher repetition rates.

It was brought to our notice that simultaneous to the acceptance of our manuscript, a 8 mJ, 1 kHz Ho:YLF regenerative amplifier was reported operating in a bistable regime [21].

**Acknowledgment**. This work was funded by the Max Planck Society. The authors thank Djordje Gitaric for his technical assistance and Haider Zia for his contribution to the Frantz-Nodvik based model.

## References

1. A. D. DiChiara, S. Ghmire, D. A. Reis, L. F. DiMauro, and P. Agostini, Strong-field and attosecond physics with mid-infrared lasers in Attosecond Physics (Springer-Verlag, 2013).
2. P. Malevich, T. Kanai, H. Hoogland, R. Holzwarth, A. Baltuska, and A. Pugzlys, CLEO: 2015, OSA Technical Digest (online) (Optical Society of America, 2015), paper SM1P.4.
3. P. Malevich, G. Andriukaitis, T. Flöry, A. Verhoef, A. Fernández, S. Ališauskas, A. Pugžlys, A. Baltuška, L. Tan, C. Chua, and P. Phua, Opt. Lett. 38, 2746-2749 (2013).
4. L. von Grafenstein, M. Bock, U. Griebner, and T. Elsaesser, 2015 European Conference on Lasers and Electro-Optics - European Quantum Electronics Conference, (Optical Society of America, 2015), paper CF_1_2.
5. A. Dergachev, Proc. SPIE 3005, 85990B (2013).
6. M. Hemmer, D. Sánchez, M. Jelínek, V. Smirnov, H. Jelinkova, V. Kubeček, and J. Biegert, Opt. Lett. 40, 451-454 (2015).
7. L. von Grafenstein, M. Bock, U. Griebner, and T. Elsaesser, Opt. Express 23, 14744-14751 (2015).
8. A. Wienke, D. Wandt, U. Morgner, J. Neumann, and D. Kracht, Opt. Express 23, 16884-16889 (2015).
9. B. M Walsh, G. W. Grew, N. P. Barnes, J. Phys. Chem. Solids 67, 1567-1582 (2006).
10. M. Schellhorn, Opt. Lett. 35, 2609-2611 (2010).
11. J.E. Swain, F. Rainer, IEEE J. Quantum Electron 5, 385-386 (1969).
12. J. Dörring, A. Killi, U. Morgner, A. Lang, M. Lederer, and D. Kopf, Opt. Express 12, 1759-1768 (2004).
13. X. D. Wang, P. Bassèras, J. Sweetser, I. A. Walmsley, and R. J. D. Miller, Opt. Lett. 15, 839-841 (1990).
14. P. Li, A. Ruehl, C. Bransley, and I. Hartl, CLEO: 2015, OSA Technical Digest (online) (Optical Society of America, 2015), paper STh1L.4.; arXiv:1509.09184.
15. L. M. Frantz and J. S. Nodvik, J. Appl. Phys. 34, 2346–2349 (1963).
16. W. Koechner, Solid-State Laser Engineering, 4th ed. (Springer-Verlag, Berlin, 1996).
17. O. J. P. Collet, MSc Thesis, "Modeling of end-pumped Ho:YLF amplifiers," University Stellenbosch (2013).
18. P. Kroetz, A. Ruehl, K. Murari, H. Cankaya, A. Calendron, F. Kaertner, I. Hartl, and R. J. D. Miller, CLEO: 2015, OSA Technical Digest (online) (Optical Society of America, 2015), paper SF1F.3.
19. H. P. Li, C. H. Kam, Y. L. Lam, W. Ji, Opt. Mater. 15, 237-242 (2000).
20. L.-T. Cheng, L.K. Cheng, J.D. Bierlein, Proc. SPIE 1863, 43-53 (1993).
21. L. von Grafenstein, M. Bock, U. Griebner, and T. Elsaesser, Advanced Solid State Lasers, OSA Technical Digest (online) (Optical Society of America, 2015), paper AW4A.8.

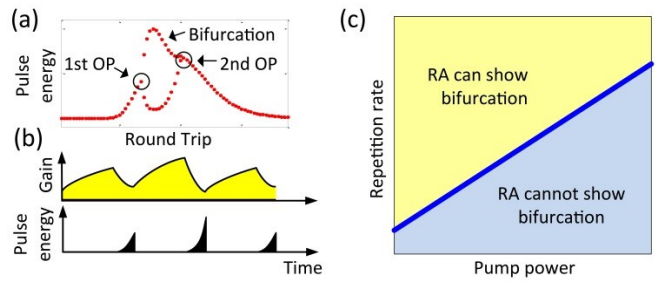

Fig. 1. (a) Schematic showing the 1st and 2nd operation points (OPs), preceding and succeeding bifurcation respectively, as a function of the number of round trips at high repetition rates [12]. (b) Schematic of the disturbed equilibrium in the gain for 3 cycles, causing bifurcation [12]. (c) Schematic showing the susceptibility of a RA to bifurcation when operated close to gain saturation. An increase in the pump power allows increased repetition rate while still suppressing bifurcation.

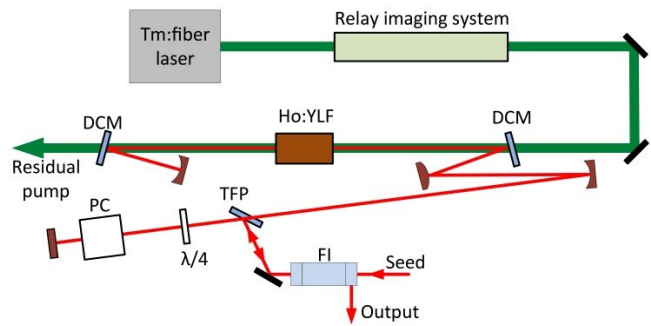

Fig 2. Schematic of the Ho:YLF regenerative amplifier system. DCM: dichroic mirror; TFP: thin film polarizer; λ/4: quarter wave-plate; PC: Pockels cell; FI: Faraday isolator.

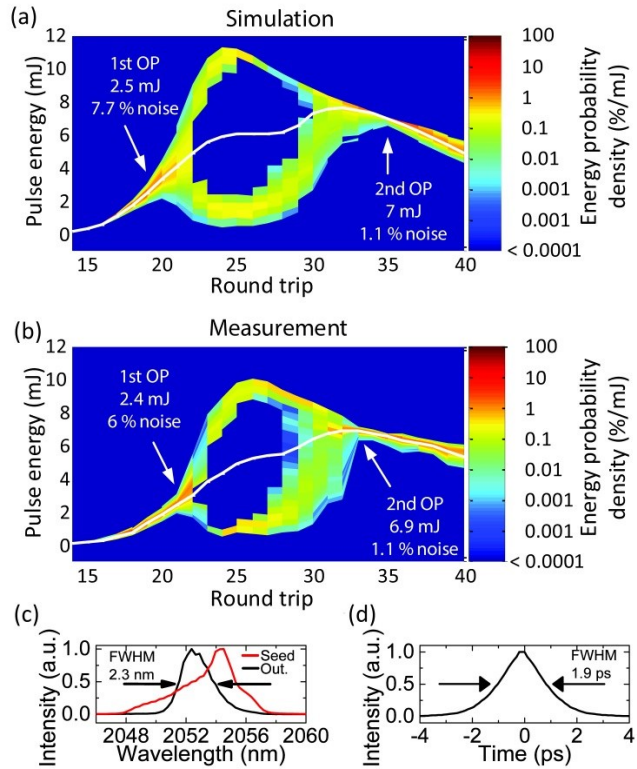

Figure 3. (a) Simulated and (b) measured energy probability density at 1 kHz repetition rate. The white line indicates the average pulse energy. Within the bifurcation regime, individual pulse energies can significantly exceed the average output pulse energy. (c) RA seed spectrum and measured RA output spectrum at 6.9 mJ. (d) Calculated Fourier-limited temporal profile corresponding to the output spectrum.

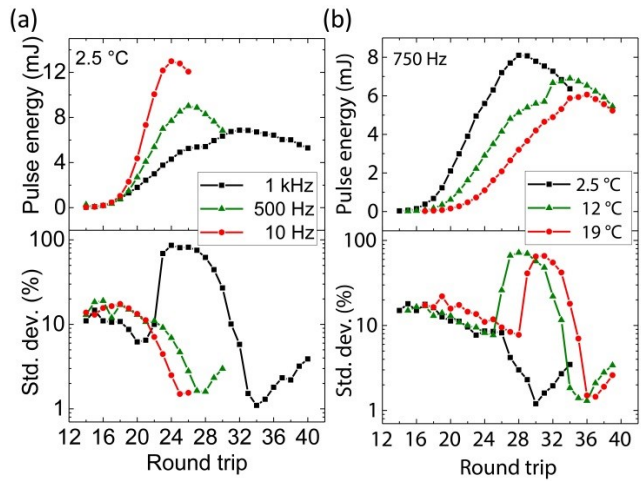

Fig 4. (a) Output pulse energies vs. round trips for the RA operated at 1 kHz, 500 Hz and 10 Hz, together with the corresponding pulse-to-pulse fluctuations. (b) Output pulse energy vs. round trip for the RA operated at 2.5 °C, 12 °C and 19 °C, together with the corresponding pulse-to-pulse fluctuations.

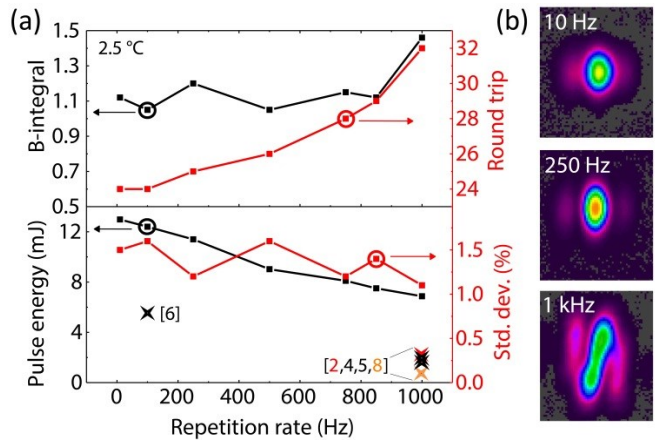

Figure 5. (a) Summary of pulse energies, noise, B-integral and number of round trips as a function of the repetition rate, together with the highest pulse energies reported in literature. Color of star indicates gain material. Red: Ho:YAG; black: Ho:YLF; orange: Tm:YAP. (b) Output beam profile at the exit of the RA for the repetition rates 10 Hz, 250 Hz and 1 kHz, at the corresponding highest pulse energy.